\documentclass[aps,showpacs,showkeys,twocolumn,nofootinbib,superscriptaddress,altaffilletter]{revtex4-2}
\usepackage{amssymb,amsmath,amsthm}
\usepackage{mathrsfs}
\usepackage{mathtools}
\usepackage[applemac]{inputenc}
\usepackage[thinc]{esdiff}
\usepackage{enumerate}
\usepackage{enumitem}
\usepackage{float}

\usepackage{changepage}%fit table

\usepackage[export]{adjustbox}

\pdfoutput=1

\usepackage[dvipsnames]{xcolor}
\usepackage{graphicx}% Include figure files
\usepackage{dcolumn}% Align table columns on decimal point
\usepackage{bm}% bold math
\usepackage[pdftex]{hyperref}% add hypertext capabilities
\hypersetup{
	colorlinks=true,
	linkcolor=black!30!blue,
	citecolor=black!30!blue,
	filecolor=black,
	urlcolor=black!45!blue,
}

\newcommand{\orcidicon}{\includegraphics[width=0.32cm]{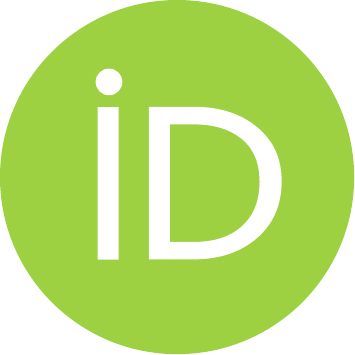}}

\newcommand\orcidT{{\href{https://orcid.org/0000-0001-8007-5181}{\orcidicon}}}
\newcommand\orcidM{{\href{https://orcid.org/0000-0002-1529-1889}{\orcidicon}}}
\newcommand\orcidP{{\href{https://orcid.org/0000-0001-8073-4896}{\orcidicon}}}

\begin{document}
	
	\title{Big rip in shift-symmetric Kinetic Gravity Braiding theories}
	
	\author{Teodor Borislavov Vasilev\,\orcidT}
	\email{teodorbo@ucm.es}
	\affiliation{Departamento de F\'isica Te\'orica and IPARCOS, Universidad Complutense de Madrid, E-28040 Madrid, Spain.}
	
	\author{Mariam Bouhmadi-L\'opez\,\orcidM}%
	\email{mariam.bouhmadi@ehu.eus}
	\affiliation{
		IKERBASQUE, Basque Foundation for Science, 48011, Bilbao, Spain.
	}%
	\affiliation{%
		Department of Physics and EHU Quantum Center, University of the Basque Country, UPV/EHU, P.O. Box 644, 48080 Bilbao, Spain.
	}

	\author{Prado Mart\'in-Moruno\,\orcidP}
	\email{pradomm@ucm.es}
	\affiliation{Departamento de F\'isica Te\'orica and  IPARCOS, Universidad Complutense de Madrid, E-28040 Madrid, Spain.}

	\date{\today}
	
	\begin{abstract}
	We revise the future fate of a group of scalar-tensor theories known as kinetic gravity braiding models. As it is well-known, these theories can safely drive the expansion of the universe towards a future de Sitter state if the corresponding Lagrangian is invariant under constant shifts in the scalar field. However, this is not the only possible future state of these shift-symmetric models as we show in this letter. In fact, future cosmic singularities characterized by a divergence of the energy density can also appear in this framework. We present an explicit example where a big rip singularity is the only possible future fate of the cosmos. 
	\end{abstract}

	\maketitle

	%%%%%%%%%%%%%%%%%%%%%%%%%%%%%%%%%%%%%%%%%%%%%%%%%%%%%%%%%%%%%%%%%%%%%%%%%%%%%%%%%%%%%%%%%%%%%%%%%%%%%%%%%%%%%%%%%%%%%%%%%%%%%%%%%%%%%%%%%%%%%%%%%%%%%%%%
	
	\section{Introduction}
	Dark energy is an unknown form of energy characterized by a negative pressure  that affects the universe on large scales. In the framework of general relativity, this exotic energy is responsible for the accelerated expansion of our universe. However, the underlying nature of this dark component is still an enigma nowadays.
	
	Fortunately, precise measurements of the equation of state $w$ for dark energy, that is the ratio between its pressure and energy density, may provide valuable hints for unveiling its mysteries. Current observations restrict this value to a narrow band around the cosmological constant case, i.e., around $w=-1$ \cite{CosmoData}. Nevertheless, from a theoretical perspective, the cosmological constant corresponds to a critical value on the $w$-line. If $w$ equals exactly $-1$, then the universe would continue expanding indefinitely and, eventually, would approach a de Sitter (dS) state. Conversely, if dark energy is characterized by $w<-1$, which is known as phantom dark energy, the universe enters in a super-accelerated regime. In this scenario, the extreme expansion rate could lead to a future cosmic singularity.
	A well-known example of such fatal fate is the big rip (BR) singularity \cite{BR,BR2} (see also \cite{NojiriClassification}).  When the universe approaches this phantom-dark-energy doomsday, all bounded structures and, ultimately, space-time itself are ripped apart as the size of the observable universe, the Hubble rate, and its cosmic time derivative diverge. Moreover, this singularity takes place in a finite cosmic time \cite{BR,BR2}. Since the possibility of having today phantom dark energy is not at all observationally excluded, revealing dark energy's nature is of capital interest for understanding the final fate of our Universe.
	
	%%%%%%%%%%%%%%%%%%%%%%%%%%%%%%%%%%%%%%%%%%%%%%%%%%%%%%%%%%%%%%%%%%%%%%%%%%%%%%%%%%%%%%%%%%%%%%%%%%%%%%%%%%%%%%%%%%%%%%%%%%%%%%%%%%%%%%%%%%%%%%%%%%%%%%%%
	
	\section{Shift-symmetric kinetic Gravity Braiding theories}
	An interesting underlying framework for dark energy models is that provided by Kinetic Gravity Braiding (KGB) theories \cite{KGB1}. The action of the KGB theory is given by \cite{KGB1}
	\begin{align}\label{KGB}
	S=\int d^4 x\sqrt{-g}\left[\frac12 R+ K(\phi, X)-G(\phi, X)\Box\phi\right],&&
	\end{align}
	where we have used the geometric unit system $8\pi G=c=1$, $K(\phi,X)$ and $G(\phi,X)$ are arbitrary functions of the scalar field and its canonical kinetic term $X\coloneqq-\frac12 g^{\mu\nu}\nabla_\mu\phi\nabla_\nu\phi$, and
	$\Box \phi=g^{\mu\nu}\nabla_\mu\nabla_\nu\phi$. A remarkably interesting case is the shift-symmetric sector of this theory. That is when the above action is symmetric under constant shifts in the scalar field, i.e. under the transformation $\phi\to\phi+c$, being $c$ a constant. In practise, this implies that the functions $K$ and $G$ must not depend on the scalar field $\phi$. In that case, the scalar field equation is given by the conservation equation of the corresponding shift-current \cite{KGB1}. 	
	In fact, the conserved shift-current provides key information about the future attractors of the theory for a homogeneous and isotropic cosmological background \cite{KGB1}. Moreover, it was also shown that the scalar field can exhibit stable phantom behaviour, i.e. free from ghost and gradient instabilities, whilst driving the expansion of the universe towards a future self-tuning (quasi)dS state \cite{KGB1} (see also references \cite{DeFelice:2010pv,Bernardo:2021hrz,Germani:2017pwt,Martin-Moruno:2015bda,Martin-Moruno:2015kaa,DeFelice:2011bh,Muharlyamov:2021dlh,Tsujikawa:2010zza}). Consequently, these models have proven to be an exceptional framework for describing extensions of the standard cosmological model. 
	Nevertheless, explicit or implicit assumptions have been usually taken  when studying the future fate of these scalar-tensor theories. Hence, limiting their phenomenology to asymptotic dS states only.
	In this letter, we provide a general discussion on how different future fates could be accommodated within the KGB framework. We also present an explicit example featuring a future BR singularity.

	We shall limit our discussion to the homogeneous and isotropic cosmological background given by the Friedmann-Lema\^{i}tre-Robertson-Walker (FLRW) line element
	\begin{eqnarray}
		ds^2=-N(t)^2dt^2+a(t)^2dx_3^2,
	\end{eqnarray}
	being $N$ the lapse function, $a$ the scale factor and $dx^2_3$ the spatial three-dimensional sections. We consider dust and radiation as external sources to action (\ref{KGB}) for the viability of the model on cosmological scales. Then, substituting this metric into the Lagrangian, a straightforward variation with respect to the lapse function and the scale factor leads to 
	\begin{align}
		3H^2=&\rho_m+\rho_r-K+\dot{\phi}J,\label{eq:FE1}\\
		\dot{H}=&-\frac12\left(\rho_m+\frac43\rho_r\right)+XG_X \ddot{\phi}-\frac12\dot{\phi}J,\label{eq:FE2}
	\end{align}
	where  $H\coloneqq\dot{a}/a$ stands for the Hubble rate, the dot represents derivation with respect to the cosmic time and
	\begin{eqnarray}\label{eq:J}
		J\coloneqq \dot{\phi}K_X+6HXG_X,
	\end{eqnarray}
	is the only non-trivial component of the shift-current \cite{KGB1}.
	In addition,  $\rho_m$ and $\rho_r$ represent the energy densities of dust and radiation, respectively.
	Their field equations read
	\begin{align}
	\dot{\rho}_m=&-3H\rho_m,\label{eq:m}\\
	\dot{\rho}_r=&-4H\rho_r.\label{eq:r}
	\end{align}
	Conversely, the evolution of the scalar field is given by the conservation equation for the shift-current (\ref{eq:J}). That is \cite{KGB1}
	\begin{eqnarray}
		\frac{1}{a^3}\diff{\left(a^3J\right)}{t}=0.\label{eq:phi}
	\end{eqnarray}	
	Hence, it is straightforward to find a first integral of motion for $\phi$. As a result
	\begin{eqnarray}\label{eq:Jsym}
		J=Q_0\left(\frac{a}{a_0}\right)^{-3},
	\end{eqnarray}
	being $Q_0$ the scalar charge associated with the shift symmetry and $a_0$ the current value of the scale factor. We recall, however, that the evolution equation (\ref{eq:phi}) for the scalar field  is not independent from equations (\ref{eq:FE1}) and (\ref{eq:FE2}). 
	
	As a result of equation (\ref{eq:Jsym}), the shift-current is either trivial, if $Q_0=0$, or tends to zero asymptotically for infinitely expanding FLRW models. 	
	Moreover, the vanishing of $J$ can be used to infer valuable information about the future attractors of the theory, as it was first noticed in reference \cite{KGB1}. Nevertheless, it should be emphasized that $J=0$ does not represent a proper fixed point of the system. Since (\ref{eq:FE1}), (\ref{eq:FE2}), (\ref{eq:m}), (\ref{eq:r}) and (\ref{eq:phi}) are four dynamical equations and one constraint, there are only three independent degrees of freedom. Therefore, the corresponding configuration space is three dimensional. Hence, the condition $J=0$ represents a surface on the phase-space. This surface either contains all the possible trajectories in the configuration space, that is if and only if $Q_0=0$, or it will be asymptotically intersected if $a$ diverges, as noted in \cite{Germani:2017pwt}.
	%
	%Note that this general result does not depend on the particular choice for the functions $K$ and $G$, but relies on the shift symmetry of the theory. 
	
	Consequently, even though the specific evolution of the system will depend on the choice for the functions $K$ and $G$, some general conclusions about the future state of the model can be deduced from equation (\ref{eq:Jsym}). If $Q_0=0$, then the Friedmann equations (\ref{eq:FE1}) and (\ref{eq:FE2}) simplify as the term $\dot{\phi}J$  drops out in virtue of equation (\ref{eq:Jsym}). Hence, the system could evolve towards a final (quasi)dS state as long as the k-essence function $K$ is asymptotically finite and negative, i.e. playing eventually the role of a positive cosmological constant, and the slow-roll-like condition $XG_X\ddot{\phi}\approx0$ is fulfilled \cite{KGB1}. Indeed, the very same is also true for $Q_0\neq0$ provided that the scalar field velocity $\dot{\phi}$ does not increase asymptotically faster than $a^3$, in which case the term $\dot{\phi}J$ will likewise vanish, eventually.
	These assumptions are usually made, either implicitly or explicitly,  in the literature when studying the future dS attractors of the shift-symmetric KGB theories \cite{KGB1} (see also, for instance, references \cite{Bernardo:2021hrz,Germani:2017pwt,Martin-Moruno:2015bda,Martin-Moruno:2015kaa,Tsujikawa:2010zza}).
	However, different future attractors are also possible. For $Q_0\neq0$, if the scalar field velocity $\dot{\phi}$ grows faster than the shrinking rate of the shift-current, then the contribution of the scalar field to the total energy budget and pressure may diverge, thus, leading to a different asymptotic state at which both the Hubble rate and its cosmic time derivate blow up. Furthermore, if that divergence occurs at a finite cosmic time, then the final fate of the model could be that of a BR singularity. Owing to equations (\ref{eq:J}) and (\ref{eq:Jsym}), for this singularity to be found the functions $K$ and $G$ should be chosen in such a way that either both terms in equation (\ref{eq:J}) cancel out each other or vanish separately as $\dot{\phi}$ and $H$ diverge. 
	(See also reference \cite{Muharlyamov:2021dlh} for a discussion of the future phenomenology when $Q_0\neq0$.)
	
	%%%%%%%%%%%%%%%%%%%%%%%%%%%%%%%%%%%%%%%%%%%%%%%%%%%%%%%%%%%%%%%%%%%%%%%%%%%%%%%%%%%%%%%%%%%%%%%%%%%%%%%%%%%%%%%%%%%%%%%%%%%%%%%%%%%%%%%%%%%%%%%%%%%%%%%%%%

	\section{Proxy model with a big rip}
	In order to present an explicit example featuring a BR singularity, we consider the simplest case still different from kinetic $k$-essence. This is obtained when only the braiding function $G$ is present in action (\ref{KGB}). For the simplicity of the discussion we assume a power law for this function; that is
	\begin{align}\label{model}
		K(X)=0 \hspace{.5cm}\textup{and}\hspace{0.5cm} G(X)=c_G X^\beta,
	\end{align}
	being $c_G$ a coupling constant and $\beta$ the parameter labelling different models. The corresponding shift-current (\ref{eq:J}) reads
	\begin{eqnarray}\label{eq:Jmodel}
		J=6\beta c_G H X^\beta.
	\end{eqnarray}
	Then, a BR event, if any, is expected to occur for some negative value of the parameter $\beta$, for which $\dot{\phi}$ and $H$ would grow limitless but $J$ would still dilute away as demanded by equation (\ref{eq:Jsym}).
	To find out whether such behaviour is present in this model, consider that there exists at least some value for $\beta$  for which the corresponding scalar field energy density does not dilute faster than matter with the expansion. Therefore, as radiation and matter are redshifted  away in the Friedmann equation (\ref{eq:FE1}), eventually the approximation
	\begin{eqnarray}\label{eq:FE1approx}
		3H^2\approx \rho_\phi=\dot{\phi}J,
	\end{eqnarray}
	holds true. Furthermore, we can also assume the scalar field effective energy density $\rho_\phi$ to be positive at least when it is the dominant component. This condition implies $\epsilon Q_0$ to be positive where $\epsilon\coloneqq\textup{sgn} \ \dot{\phi}$; see equations (\ref{eq:Jsym}) and (\ref{eq:FE1approx}). [Also note that $Q_0$ should have the same sign as the product $\beta c_G H$ according to (\ref{eq:Jsym}) and (\ref{eq:Jmodel}).]
	On the other hand, comparing equations (\ref{eq:Jsym}) and (\ref{eq:Jmodel}), the canonical kinetic term $X$ can be expressed as a function on the Hubble rate and the scale factor, that is $X=X\left(a,H;Q_0\right)$. Thus, the approximation (\ref{eq:FE1approx}) leads to 
	\begin{eqnarray}
		H\approx\lambda \left(\frac{a}{a_0}\right)^{-3\frac{2\beta+1}{4\beta+1}},
	\end{eqnarray} 
	being $\lambda\coloneqq(\sqrt{2}\epsilon Q_0/3)^\frac{2\beta}{4\beta+1}\left(Q_0/6\beta c_G\right)^{\frac{1}{4\beta+1}}$ a positive constant for an expanding universe $(H>0)$. Then, when $\beta$ is less than $-1/4$ the scalar field contribution does not dilute faster than matter.
	In view of equation (\ref{eq:FE1approx}), if $\beta=-1/2$ the model evolves towards a final dS state given by $H_\textup{dS}=-\sqrt{2}\epsilon c_{G}$ (where $H_\textup{dS}>0$). However, for $\beta\in(-1/2,-1/4)$ the exponent in the preceding equation becomes positive and, therefore, the value of the Hubble rate increases with the scale factor. It is a well-known fact that in latter case, i.e. whenever the Hubble rate is proportional to a positive power of the scale factor $a$, both $H$ and $\dot{H}$ blow up in a finite cosmic time \cite{BR,BR2}. This behaviour corresponds to a genuine BR singularity. Therefore, when $\beta\in(-1/2,-1/4)$ the future fate of the model is that of a BR singularity.
	
	\begin{table*}[t]	
		\setlength\arrayrulewidth{0.6pt}
		\renewcommand{\arraystretch}{1.5}
		%%%%%%%%%%
		\begin{adjustwidth}{-.35cm}{}
		\begin{tabular}{l c c c c c c c c c c c}
			\hline \hline 
			Fixed Point & $(h^{\textup{fp}},\Omega_\phi^{\textup{fp}},\Omega_r^{\textup{fp}})$ & $w_\phi^{\textup{fp}}$ & $w_{\textup{eff}}^{\textup{fp}}$ & $\beta<-\frac12$ & $\beta=-\frac12$ & $-\frac12<\beta<-\frac14$ &
			$\beta=-\frac14$ & $-\frac14<\beta<0$ & $0<\beta<\frac12$ & $\beta=\frac12$ & $\frac12<\beta$  \\\hline
			$\textup{P}_1$ (vacuum)& $(0,0,0)$ & $\frac{1}{4\beta}$ & $0$ & saddle  & saddle & saddle & saddle & saddle & attractor & attractor &  attractor \\ 
			$\textup{P}_2$ (vacuum)& $(0,1,0)$ & $\frac{1}{4\beta+1}$ & $\frac{1}{4\beta+1}$ & attractor  & | & saddle & | & saddle & saddle & | & saddle \\ 
			$\textup{P}_3$ (vacuum)& $(0,0,1)$ & $\frac{1}{6\beta}$ & $\frac{1}{3}$ & saddle  & saddle & saddle & saddle & saddle & saddle & | &  saddle \\ 
			$\textup{P}_4$ (BB)& $(1,0,0)$ & $\frac{1}{4\beta}$ & 0 & saddle  & saddle & saddle & saddle & saddle & saddle & saddle & saddle \\ 
			$\textup{P}_5$ (BB/BR)& $(1,1,0)$ & $\frac{1}{4\beta+1}$ &  $\frac{1}{4\beta+1}$ & saddle  & | & attractor & | & repeller & repeller & | & saddle \\ 
			$\textup{P}_6$ (BB)& $(1,0,1)$ & $\frac{1}{6\beta}$ & $\frac{1}{3}$ & repeller  & repeller & repeller & repeller & repeller & saddle & | & repeller \\ 
			$\textup{P}_7$ (BF)& $(1,1,0)$ & $-\infty$ & $-\infty$ & |  & | & | & attractor$^\star$ & | & | & | & | \\
			${\textup{L}_1}$(dS) & $(h^\textup{fp},1,0)$ & $-1$ & $-1$ & | & attractor & | & |  & |  & |  & |  & |  \\
			${\textup{L}_2}$ (sudden)& $(h^\textup{fp},-4\beta,\Omega_r^{\textup{fp}})$ & $-\infty$ & $-\infty$ & | & | & | & |  &  attractor$^\star$  & | & |  & |  \\ 
			$\textup{L}_3$ (vacuum)& $(0,\Omega_\phi^{\textup{fp}},\Omega_r^{\textup{fp}})$ & $\frac{1}{3}$ & $\frac{1}{3}$ & | & | & | & | & | & | & saddle & | \\ 
			$\textup{L}_4$ (BB)& $(1,\Omega_\phi^{\textup{fp}},\Omega_r^{\textup{fp}})$ & $\frac{1}{3}$ & $\frac{1}{3}$ & | & | & | & | & | & | & repeller & | \\
			\hline 
		\end{tabular}
		\end{adjustwidth}
		\caption{Classification and linear stability of the fixed points of our model. A superscript ``fp'' indicates evaluation at the fixed point. A horizontal bar denotes that the corresponding fixed point does not exist. The physical interpretation of each point is shown in brackets where BB stands for Big Bang and BF for Big Freeze. The labels L$_1$, L$_2$, L$_3$ and L$_4$ represent sets of non-isolated fixed points where $h^{\textup{fp}}$ can take any values. In addition,  $\Omega_r^{\textup{fp}}\in[0,1+4\beta]$ holds for L$_2$, and  $\Omega_\phi^{\textup{fp}}+\Omega_r^{\textup{fp}}=1$ for L$_3$ and L$_4$. The starred quantities designate fixed points that have eluded our dynamical system analysis because of the choice of the dynamical variables but whose existence and stability follows directly from the Friedmann equations.\label{tab:FPs}}
	\end{table*}
	
	The presence of a future BR singularity in this model is also confirmed with a dynamical system analysis. 
	In view of equation (\ref{eq:FE1}), we define the dimensionless variables
	\begin{align}
		\Omega_r&\coloneqq \frac{\rho_r}{3H^2},\label{varOR}\\
		\Omega_m&\coloneqq \frac{\rho_m}{3H^2},\\
		\Omega_\phi&\coloneqq \frac{\dot{\phi} J}{3H^2}\label{varOP},
	\end{align}
	where for the sake of the argumentation $\Omega_\phi$ is assumed to be positive, which is equivalent to considering expanding FLRW only. Hence, $\Omega_i\in[0,1]$ for $i=\left\lbrace m,r,\phi\right\rbrace$. (Also recall that $K$ is trivially zero for the model at hands.) With these definitions, the Friedmann equation (\ref{eq:FE1}) can be re-written as
	\begin{eqnarray}\label{eq:FE1omegas}
		\Omega_r+\Omega_m+\Omega_\phi=1.
	\end{eqnarray}
	This constraint allows the elimination of one of the aforementioned variables from the dynamical system. Then, an autonomous system is obtained with the introduction of a suitable fourth variable. In order to obtain a compact system we select the new variable, $h$, as the following compactification for the Hubble rate \cite{FPatInfty}	
	\begin{align}\label{varh}
		\frac{h}{1-h^2}=\frac{H}{H_0},
	\end{align}
	with $H_0$ the current value of the Hubble rate and $h\in[0,1]$  since we focuss only in expanding geometries.
	It should be emphasized that compact variables are highly recommended, otherwise fixed points at the infinite boundary of the system may be overlooked. Using these new variables, the evolution equations for the cosmological for our model are
	\begin{align}
		h'&=-\frac{(1-h^2)h\left(2\beta\Omega_r+6\beta+3\Omega_\phi\right)}{\left(1+h^2\right)\left(4\beta+\Omega_\phi\right)},\label{eq:Ph}\\
		\Omega_\phi'&=\Omega_\phi\frac{(1+4\beta)\Omega_r+3\left(\Omega_\phi-1\right)}{4\beta+\Omega_\phi},\label{eq:POP}\\
		\Omega_r'&=2\Omega_r\frac{2\beta\left(1-\Omega_r\right)-\Omega_\phi}{4\beta+\Omega_\phi},\label{eq:POR}
		%&\Omega_m=1-\Omega_r-\Omega_\phi,\label{eq:POM}
	\end{align}
	where a prime denotes differentiation with respect to the dimensionless time-like variable\footnote{This definition for the independent time variable of the system is only valid for monotonically expanding geometries. Therefore, contracting universes or turnaround/bounce-like events, if any, cannot be described within this dynamical system formulation.} $x\coloneqq\ln (a/a_0)$. In addition, the effective equation of state parameter and the equation of state parameter for the scalar field are	
	\begin{align}
		w_{\textup{eff}}&\coloneqq \frac{P_{\textup{total}}}{\rho_{\textup{total}}}=-1+2\frac{2\beta(3+\Omega_r)+3\Omega_\phi}{3(4\beta+\Omega_\phi)},\\
		w_\phi&\coloneqq \frac{P_\phi}{\rho_\phi}=-1-\frac{\Omega_r-3(1+4\beta+\Omega_\phi)}{3(4\beta+\Omega_\phi)},
	\end{align}
	respectively, where $P_{\textup{total}}$, $\rho_{\textup{total}}$, $P_{\phi}$ and $\rho_{\phi}$ are taken from the r.h.s. of equations (\ref{eq:FE1}) and (\ref{eq:FE2}).
	
	The fixed points of the dynamical system (\ref{eq:Ph})-(\ref{eq:POR}) are listed in table \ref{tab:FPs}. Recall that the physical meaning of these points showed in table \ref{tab:FPs} has been deduced from the perspective of an expanding universe only. Along this line, P$_4$ has been classified as a matter dominated initial Big Bang (BB) singularity where only some trajectories (those where radiation is absent) may begin in P$_4$ if matter dominates over the scalar field when $a\to 0 $.
	
	Note that the case of $\beta=0$ is not presented in table \ref{tab:FPs} since it corresponds to a universe filled solely with dust and radiation. Also note that the aforementioned dynamical system is ill-defined when $\Omega_\phi=-4\beta$, which occurs in the physical configuration space whenever $\beta\in[-1/4,0)$. In fact, direct evaluation of the Friedmann equations (\ref{eq:FE1}) and (\ref{eq:FE2}) reveals that the system always evolves towards $\Omega_\phi=-4\beta$ for that range of values for $\beta$. Therefore, this behaviour indeed constitutes an attractor in the model, although the choice of variables in (\ref{varOR})-(\ref{varOP}) and (\ref{varh}) is not adequate for its analysis. (The physical interpretation of these attractors is briefly depicted in table \ref{tab:FPs} and it will be properly addressed in a future work.)
	
	Note that there are only one repeller and one attractor when $\beta\in(-1/2,-1/4)$, namely $\textup{P}_6$ and $\textup{P}_5$, respectively. The former, characterized by the divergence of the Hubble rate in a radiation-dominated epoch, corresponds to a BB singularity. The later, given by the divergence of the Hubble rate in a phantom-scalar-field-dominated era ($w_{\textup{eff}}=w_{\phi}<-1$), is precisely the future BR singularity discussed before. Hence, the analysis of the fixed points of the system (\ref{eq:Ph})-(\ref{eq:POR}) confirms the presence of a future BR when $\beta\in(-1/2,-1/4)$, being that fatal fate de facto the only future attractor in the corresponding configuration space.
	Further details on the dynamical system description of future phantom attractors and the analysis of different shift-symmetric KGB models will be addressed in an incoming work. 
	%%%%%%%
	
\begin{figure}[t]
		\includegraphics[width=\columnwidth]{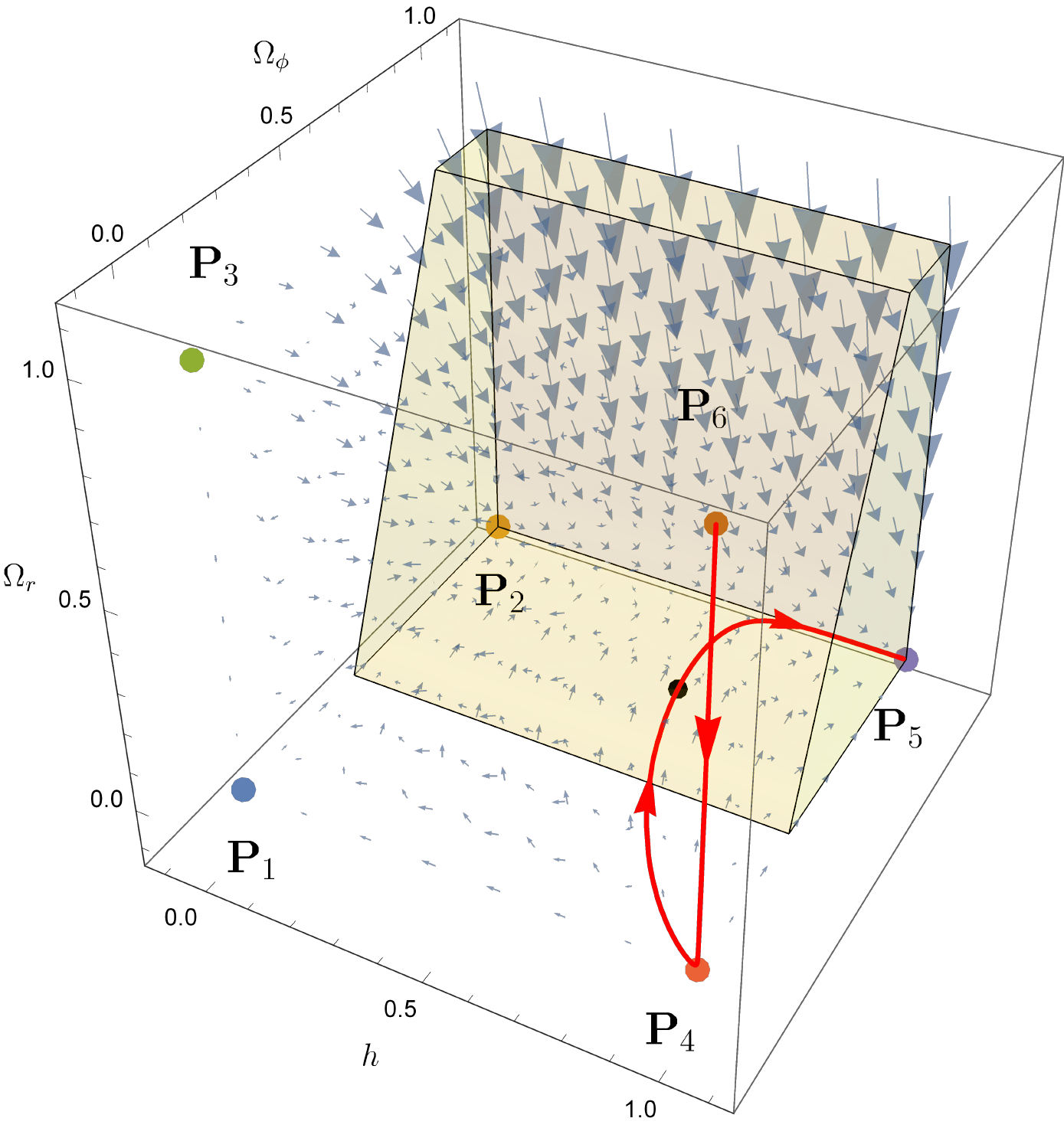}
		\caption{Phase-space portrait of the  dynamical system (\ref{eq:Ph})-(\ref{eq:POR}) for $\beta=-2/5$. The red trajectory was drawn for the physical values $H_0=67.27$ km s$^{-1}$ Mpc$^{-1}$, $\Omega_{\phi0}=0.6836$, $\Omega_{m0}=0.3164$ and $\Omega_{r0}=8.4\times10^{-5}$ \cite{CosmoData}. The black dot on the trajectory corresponds to the present state of the system. The yellow-shaded volume represents the region in the configuration space where the universe is accelerating.}
		\label{fig:flow}		
	\end{figure}

\begin{figure}[t]
		\includegraphics[width=\columnwidth]{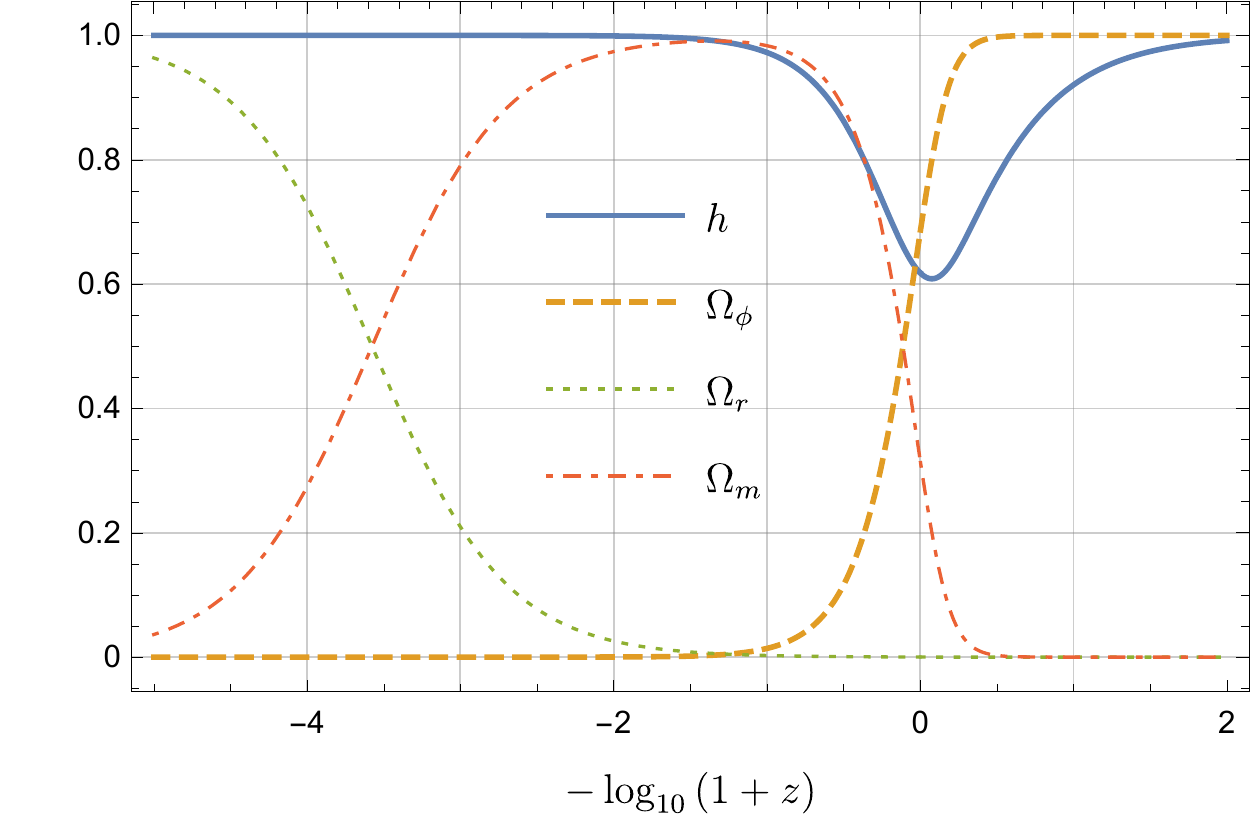}
		\includegraphics[width=1.016\columnwidth,right]{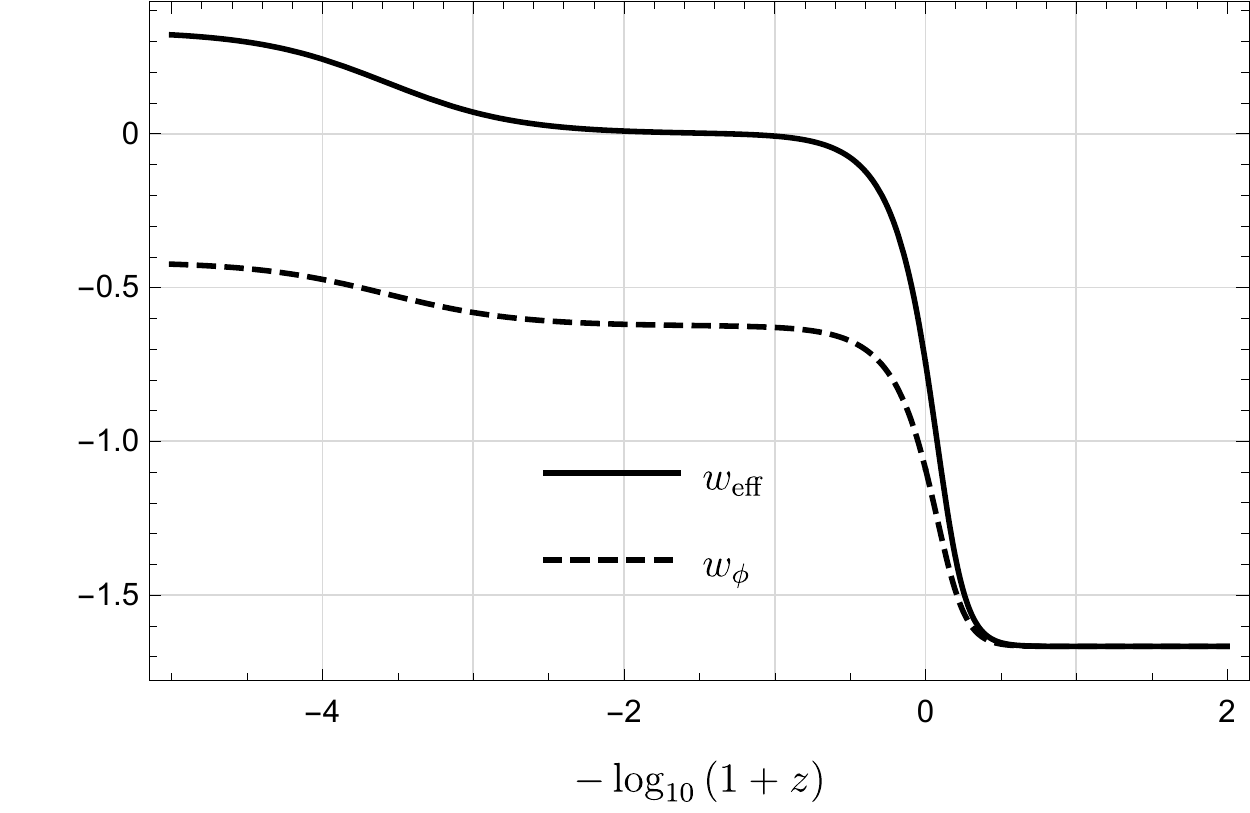}
		\caption{Numerical evolution of the dynamical system (\ref{eq:Ph})-(\ref{eq:POR}) for
			$\beta=-2/5$ with the same initial conditions as in figure \ref{fig:flow}. 
			Top panel: the variable $h$ and the partial densities $\Omega_i$
			for $i=\left\lbrace m,r,\phi\right\rbrace$.
			Bottom panel: the effective equation of state parameter  $w_\textup{eff}$ and the equation of state parameter $w_\phi$ for the scalar field. \label{fig:evolution}}		
	\end{figure}

	For the sake of the discussion, the configuration space for the dynamical system (\ref{eq:Ph})-(\ref{eq:POR}) with $\beta=-2/5$ is represented in figure \ref{fig:flow}. The initial conditions $h_0=(\sqrt{5}-1)/2\approx0.618$, $\Omega_{m0}=0.3164$, $\Omega_{r0}=8.4\times10^{-5}$ and $\Omega_{\phi0}=1-\Omega_{m0}-\Omega_{r0}$ have been taken at face value \cite{CosmoData} for the numerical integration. The corresponding trajectory in the phase-space approaches the repulsive fixed point P$_6$ in the asymptotic past of the system.
	Since radiation dominates over matter at small scale factor this system would never pass through P$_4$. (In fact, the apparent proximity of the trajectory to P$_4$ is but a visual artefact of the compactification in the Hubble rate.) 
	The system eventually reaches the only attractor in the phase-space for an expanding universe, i.e. the equilibrium point P$_5$.
	Figure \ref{fig:evolution} provides further information about the cosmic history of the model.	
	Owing to 
	the evolution of the equation of state parameter of the total fluid $w_\textup{eff}$ in figure \ref{fig:evolution}, the universe has entered the accelerated expansion phase at roughly $z\sim0.46$. The transition from matter to scalar field dominance has occurred at $z\sim 0.29$. Moreover, the scalar field has recently become phantom ($w_\phi<-1$) at redshift $z\sim0.12$; however, the total fluid in the r.h.s. of equation (\ref{eq:FE1}) will not exhibit effective phantom behaviour ($w_\textup{eff}<-1$) until $z\sim-0.16$, moment at which $\dot{H}$ will change its sign and, therefore, the Hubble rate will become an increasing function on time.  Accordingly, this model will meet a BR singularity in approximately 21 Gyr from present epoch.
	
	Note that the (past) expansion history of this simple model is quite compelling at the background level.
	Nevertheless, we recall that figure \ref{fig:evolution} does not represent a proper fit of the model with cosmological data but a  qualitative analysis of the expansion history of the theory where data from reference \cite{CosmoData} have been taken for the numerical integration. We expect a proper confrontation with observational data to improve the cosmological features of the model without a qualitative change in its behaviour. 
	

	%%%%%%%%%%%%%%%%%%%%%%%%%%%%%%%%%%%%%%%%%%%%%%%%%%%%%%%%%%%%%%%%%%%%%%%%%%%%%%%%%%%%%%%%%%%%%%%%%%%%%%%%%%%%%%%%%%%%%%%%%%%%%%%%%%%%%%%%%%%%%%%%%%%%%%%%%%

	\section{Concluding remarks}
	The possibility of a stable and self-tuning evolution towards a future dS state in the framework of scalar-tensor theories has naturally attracted the attention of the scientific community \cite{KGB1,Bernardo:2021hrz,Germani:2017pwt,Martin-Moruno:2015bda,Martin-Moruno:2015kaa}. Furthermore, revising the literature it could seem that this is the only possible future evolution for the shift-symmetric KGB cosmological models. Nevertheless, in this letter we have proven that different future fates are also possible. More precisely, we have provided an example of a shift-symmetric KGB model featuring a future BR singularity. This is, up to our knowledge, the first time this event has been explicitly found in the shift-symmetric sector of KGB theories. Moreover, we have also presented a brief discussion on its cosmological evolution, showing that this simple model has indeed a compelling behaviour at the background level. 
	 
	At the level of cosmological perturbations, the conditions for the absence of ghost and gradient instabilities associated with scalar, vector and tensor perturbations were first derived for the KGB theory in references \cite{KGB1,DeFelice:2011bh}. Unfortunately, the conditions there discussed for healthy scalar perturbations are not fulfilled by the BR solutions found for the model presented in equation (\ref{model}); see table \ref{tab:FPs}.  
	Nevertheless, it would be interesting to investigate whether the braiding term could introduce non-adiabatic perturbations at the linear level in the perturbation theory. This could be possible due to the presence of the Hubble rate in the energy density of the scalar field. Hence, if that is the case, it would be worthwhile to explore whether the non-adiabatic regime could ease the instabilities of the model; see reference \cite{Albarran:2016mdu} for a similar discussion for a phantom DE model with a future BR singularity.
	
	Finally, it should also be emphasised that this model is a very simple example useful for the demonstration on how future cosmic singularities can be accommodate within the shift-symmetric KGB framework. We commit the study of more elaborated models to a future work.
	
	%%%%%%%%%%%%%%%%%%%%%%%%%%%%%%%%%%%%%%%%%%%%%%%%%%%%%%%%%%%%%%%%%%%%%%%%%%%%%%%%%%%%%%%%%%%%%%%%%%%%%%%%%%%%%%%%%%%%%%%%%%%%%%%%%%%%%%%%%%%%%%%%%%%%%%%%%%
	%\vspace{.2cm}
	\section*{Acknowledgments}
	
	The research of T.B.V. and P.M.M. is supported by MINECO (Spain) Project No. PID2019-107394GB-I00 (AEI/FEDER, UE). 
	T.B.V. also acknowledges financial support from Universidad Complutense de Madrid and Banco de Santander through Grant No. CT63/19-CT64/19. He is also grateful for the hospitality of the University of the Basque country (UPV-EHU) where this work was partly developed.
	The work of M.B.L. is supported by Ikerbasque. She also would like to acknowledge the financial support from the Basque government Grant No. IT1628-22 (Spain). Her work has been supported as well by the Spanish project PID2020-114035GB-100 (MINECO/AEI/FEDER,	UE).
	%%%%%%%%%%%%%%%%%%%%%%%%%%%%%%%%%%%%%%%%%%%%%%%%%%%%%%%%%%%%%%%%%%%%%%%%%%%%%%%%%%%%%%%%%%%%%%%%%%%%%%%%%%%%%%%%%%%%%%%%%%%%%%%%%%%%%%%%%%%%%%%%%%%%%%%%%%
	
	%\appendix

	%%%%%%%%%%%%%%%%%%%%%%%%%%%%%%%%%%%%%%%%%%%%%%%%%%%%%%%%%%%%%%%%%%%%%%%%%%%%%%%%%%%%%%%%%%%%%%%%%%%%%%%%%%%%%%%%%%%%%%%%%%%%%%%%%%%%%%%%%%%%%%%%%%%%%%%%%%%%%%%%%


\begin{thebibliography}{X}
		
		%%%%%%%%%%%%%%%%%%%%%%%%%%%%%%%%%%%%%%%%%%%%%%% Observational data
	
	\bibitem{CosmoData}
	N.~Aghanim \textit{et al.} [Planck], 
	``Planck 2018 results. VI. Cosmological parameters'', 
	Astron. Astrophys. \textbf{641} (2020), A6,
	%doi:10.1051/0004-6361/201833910
	[\href{https://arxiv.org/abs/1807.06209}{arXiv:1807.06209 [astro-ph.CO]}].
	
	%%%%%%%%%%%%%%%%%%%%%%%%%%%%%%%%%%%%%%%%%%%%%%%%%%% BR
	
	\bibitem{BR}
	A.~A.~Starobinsky,
	``Future and origin of our universe: Modern view'',
	Grav. Cosmol. \textbf{6} (2000), 157-163,
	[\href{https://arxiv.org/abs/astro-ph/9912054}{arXiv:astro-ph/9912054 [astro-ph]}].
	
	\bibitem{BR2}
	R.~R.~Caldwell, M.~Kamionkowski and N.~N.~Weinberg, 
	``Phantom Energy: Dark Energy with $w<\ensuremath{-}1$ Causes a Cosmic Doomsday'', 
	Phys. Rev. Lett. \textbf{91} (2003), 071301,
	%doi:10.1103/PhysRevLett.91.071301
	[\href{https://arxiv.org/abs/astro-ph/0302506}{arXiv:astro-ph/0302506 [astro-ph]}].
	
	\bibitem{NojiriClassification}
	S.~Nojiri, S.~D.~Odintsov and S.~Tsujikawa, ``Properties of singularities in (phantom) dark energy universe'', Phys. Rev. D \textbf{71} (2005) 063004,
	%doi:10.1103/PhysRevD.71.063004
	[\href{https://arxiv.org/abs/hep-th/0501025}{arXiv:hep-th/0501025 [hep-th]}].
	
	%%%%%%%%%%%%%%%%%%%%%%%%%%%%%%%%%%%%%%%%%%%%%%%%% References on KGB
	
	\bibitem{KGB1}
	C.~Deffayet, O.~Pujolas, I.~Sawicki and A.~Vikman, 
	``Imperfect Dark Energy from Kinetic Gravity Braiding'',
	JCAP \textbf{10} (2010), 026,
	%doi:10.1088/1475-7516/2010/10/026
	[\href{https://arxiv.org/abs/1008.0048}{arXiv:1008.0048 [hep-th]}].
	
	\bibitem{Tsujikawa:2010zza}
	S.~Tsujikawa, 
	``Modified gravity models of dark energy'',
	Lect. Notes Phys. \textbf{800} (2010), 99-145,
	%doi:10.1007/978-3-642-10598-2\_3
	[\href{https://arxiv.org/abs/1101.0191}{arXiv:1101.0191 [gr-qc]}].
	
	\bibitem{DeFelice:2010pv}
	A.~De Felice and S.~Tsujikawa, 
	``Cosmology of a covariant Galileon field'', 
	Phys. Rev. Lett. \textbf{105} (2010), 111301,
	%doi:10.1103/PhysRevLett.105.111301
	[\href{https://arxiv.org/abs/1007.2700}{arXiv:1007.2700 [astro-ph.CO]}].
	
	\bibitem{DeFelice:2011bh}
	A.~De Felice and S.~Tsujikawa, 
	``Conditions for the cosmological viability of the most general scalar-tensor theories and their applications to extended Galileon dark energy models'',
	JCAP \textbf{02} (2012), 007,
	%doi:10.1088/1475-7516/2012/02/007
	[\href{https://arxiv.org/abs/1110.3878}{arXiv:1110.3878 [gr-qc]}].
	
	\bibitem{Martin-Moruno:2015bda}
	P.~Mart\'in-Moruno, N.~J.~Nunes and F.~S.~N.~Lobo,
	``Horndeski theories self-tuning to a de Sitter vacuum'', 
	Phys. Rev. D \textbf{91} (2015) no.8, 084029,
	%doi:10.1103/PhysRevD.91.084029
	[\href{https://arxiv.org/abs/1502.03236}{arXiv:1502.03236 [gr-qc]}].
	
	\bibitem{Martin-Moruno:2015kaa}
	P.~Mart\'in-Moruno and N.~J.~Nunes, 
	``Attracted to de Sitter II: cosmology of the shift-symmetric Horndeski models'', 
	JCAP \textbf{09} (2015), 056,
	%doi:10.1088/1475-7516/2015/09/056
	[\href{https://arxiv.org/abs/1506.02497}{arXiv:1506.02497 [gr-qc]}].
	
	\bibitem{Germani:2017pwt}
	C.~Germani and P.~Mart\'in-Moruno, 
	``Tracking our Universe to de Sitter by a Horndeski scalar'', 
	Phys. Dark Univ. \textbf{18} (2017), 1-5,
	%doi:10.1016/j.dark.2017.09.002
	[\href{https://arxiv.org/abs/1707.03741}{arXiv:1707.03741 [gr-qc]}].
	
	\bibitem{Bernardo:2021hrz}
	R.~C.~Bernardo, 
	``Self-tuning kinetic gravity braiding: Cosmological dynamics, shift symmetry, and the tadpole'',
	JCAP \textbf{03} (2021), 079,
	%doi:10.1088/1475-7516/2021/03/079
	[\href{https://arxiv.org/abs/2101.00965}{arXiv:2101.00965 [gr-qc]}].
	
	\bibitem{Muharlyamov:2021dlh}
	R.~K.~Muharlyamov and T.~N.~Pankratyeva, 
	``Reconstruction method in the kinetic gravity braiding theory with shift-symmetric'',	
	Eur. Phys. J. Plus \textbf{136} (2021) no.5, 590,
	%doi:10.1140/epjp/s13360-021-01607-5
	[\href{https://arxiv.org/abs/2110.15396}{arXiv:2110.15396 [gr-qc]}].
	
	%%%%%%%%%%%%%%%%%%%%%%%%%%%%%%%%%%%%%%%%%%%%% Infinity in dynamical systems
	
	\bibitem{FPatInfty}
	M.~Bouhmadi-L\'opez, J.~Marto, J.~Morais and C.~M.~Silva, 
	``Cosmic infinity: A dynamical system approach'',
	JCAP \textbf{03} (2017), 042,
	%doi:10.1088/1475-7516/2017/03/042
	[\href{https://arxiv.org/abs/1611.03100v2}{arXiv:1611.03100 [gr-qc]}].	
	
	%%%%%%%%%%%%%%%%%%%%%%%%%%%%%%%%%%%%%%%%%%%%%% Example non-adiabatic pertubations
	
	\bibitem{Albarran:2016mdu}
	I.~Albarran, M.~Bouhmadi-L\'opez and J.~Morais, 
	``Cosmological perturbations in an effective and genuinely phantom dark energy Universe'',	
	Phys. Dark Univ. \textbf{16} (2017), 94-108,
	%doi:10.1016/j.dark.2017.04.002
	[\href{https://arxiv.org/abs/1611.00392v3}{arXiv:1611.00392 [astro-ph.CO]}].
	
	
	
	
	
	
	

	
	\end{thebibliography}
\end{document}